\documentclass[12pt]{article}%
\usepackage[nosort]{cite}
\usepackage[usenames, dvipsnames]{xcolor}
\usepackage{graphicx}
\usepackage{multicol}
\usepackage{amsfonts}
\usepackage{amssymb}
\usepackage{amsmath}
\usepackage{heck}
\usepackage{afterpage}
\usepackage{setspace}
\usepackage{verbatim}
\usepackage{color}
\usepackage{longtable}
\usepackage{float}
\usepackage{subcaption}
\usepackage{epsfig}
\usepackage{enumerate}
\usepackage{epstopdf}
\usepackage[enableskew, vcentermath]{youngtab}
\usepackage{adjustbox}
\usepackage{multirow}
\usepackage{tikz}
\usepackage[margin=1in]{geometry}
\usepackage{titletoc}
\usepackage{hyperref}
\usepackage[percent]{overpic}
\usepackage{gensymb}
\usepackage{mathtools}%
\usepackage{tikz-cd}
\providecommand{\U}[1]{\protect\rule{.1in}{.1in}}
\pdfoutput=1
\newsavebox{\mysavebox}

\hypersetup{colorlinks,citecolor=black,filecolor=black,linkcolor=black,urlcolor=black}
\usetikzlibrary{decorations.markings}

\numberwithin{equation}{section}

\usetikzlibrary{chains}
\allowdisplaybreaks
\tikzset{node distance=2em, ch/.style={circle,draw,on chain,inner sep=2pt},chj/.style={ch,join},every path/.style={shorten >=4pt,shorten <=4pt},line width=1pt,baseline=-1ex}

\newcommand{\ba}{\begin{eqnarray}}
\newcommand{\ea}{\end{eqnarray}}

\newcommand{\be}{\begin{equation}}
\newcommand{\ee}{\end{equation}}

\tikzstyle{startstop} = [rectangle, rounded corners, minimum width=3cm, minimum height=1cm,text centered, draw=black, fill=blue!10]
\tikzstyle{startstop} = [rectangle, rounded corners, minimum width=3cm, minimum height=1cm,text centered, draw=black, fill=blue!10]
\tikzstyle{io} = [trapezium, trapezium left angle=70, trapezium right angle=110, minimum width=3cm, minimum height=1cm, text centered, draw=black, fill=blue!30]
\tikzstyle{process} = [rectangle, minimum width=3cm, minimum height=1cm, text centered, draw=black, fill=orange!30]
\tikzstyle{decision} = [diamond, minimum width=3cm, minimum height=1cm, text centered, draw=black, fill=green!30]
\tikzstyle{arrow} = [thick,->,>=stealth]
\tikzset{->-/.style={decoration={
  markings,
  mark=at position #1 with {\arrow[scale=2.4]{>}}},postaction={decorate}}}
\makeatletter \@addtoreset{equation}{section} \makeatother

\begin{document}

\date{December 2020}

\title{On the Swampland Cobordism Conjecture \\[4mm] and Non-Abelian Duality Groups}

\institution{PENN}{\centerline{Department of Physics and Astronomy, University of Pennsylvania, Philadelphia, PA 19104, USA}}

\authors{Markus Dierigl\footnote{e-mail: {\tt markusd@sas.upenn.edu}} and Jonathan J.\ Heckman\footnote{e-mail: {\tt jheckman@sas.upenn.edu}}}

\abstract{\noindent 
We study the cobordism conjecture of McNamara and Vafa which asserts that the bordism group of quantum gravity is trivial. In the context of type IIB string theory compactified on a circle, this predicts the presence of D7-branes. On the other hand, the non-Abelian structure of the IIB duality group $SL(2,\mathbb{Z})$ implies the existence of additional $[p,q]$ 7-branes. We find that this additional information is instead captured by the space of closed paths on the moduli space of elliptic curves parameterizing distinct values of the type IIB axio-dilaton. This description allows to recover the full structure of non-Abelian braid statistics for 7-branes. Combining the cobordism conjecture with an earlier Swampland conjecture by Ooguri and Vafa, we argue that only certain congruence subgroups $\Gamma \subset SL(2,\mathbb{Z})$ specifying genus zero modular curves can appear in 8D F-theory vacua. This leads to a successful prediction for
the allowed Mordell--Weil torsion groups for 8D F-theory vacua.
}

\maketitle

\setcounter{tocdepth}{2}

\tableofcontents

\enlargethispage{\baselineskip}

\newpage

\section{Introduction}
\label{sec:INTRO}

The central question of high energy theory revolves around understanding the unification of quantum mechanics and gravity.
From the standpoint of effective field theory, coupling a quantum field theory to gravity appears to impose seemingly mild conditions
on the structure of higher dimension operators. On the other hand, explicit string constructions always appear to impose a far greater
amount of structure on the low energy effective field theory, suggesting that present bottom up considerations provide only a partial understanding of UV consistency.

In recent years these conditions have been systematized into a set of interconnected ``Swampland Conjectures'' \cite{Vafa:2005ui, Ooguri:2006in} (see also \cite{Brennan:2017rbf, Palti:2019pca}). In fact, for supersymmetric theories in ten dimensions, string theory is universal \cite{Adams:2010zy, Kim:2019vuc}, and there is some hope that this notion of string universality can be extended to lower-dimensional systems (see e.g. \cite{Garcia-Etxebarria:2017crf, Kim:2019vuc, Cvetic:2020kuw, Ooguri:2020sua, Montero:2020icj}).

Our starting assumption, as in \cite{McNamara:2019rup} (see also \cite{Ooguri:2020sua, Montero:2020icj}), is that the underlying theory of quantum gravity is unique. This should be understood as the statement that there are finite energy defects that can interpolate between all possible vacua after a full compactification of the theory. Often this notion of connectedness demands the introduction of new ingredients in the theory which can be identified with physical objects. In particular, this is the case in situations where the bordism group of a bundle is non-trivial. The defects then trivialize this Abelian group and thus eliminate the appearance of corresponding global symmetries that would render the theory inconsistent.

In reference \cite{McNamara:2019rup} this was formalized as the conjecture that the bordism group of quantum gravity is trivial, namely
$\Omega^{\text{QG}} = 0$. One can extend this to reference a quantum theory of gravity with $D$ macroscopic dimensions $\Omega^{\text{QG},D}$,
as well as compactifications to $D- k$ large dimensions by taking $k$ of them to be small. In this case, there is a
standard notion of a bordism group $\Omega_{k}$, with $k+1$ denoting the dimension of the ``bulk manifold'' for a $k$ dimensional boundary.
While arguments involving black objects and global symmetries strictly only apply for $k \geq 3$, it is still believed that all the above groups have to vanish.\footnote{Note also that there are general caveats for Swampland criteria in theories with $D \leq 3$ spacetime dimensions since in these cases there are no local degrees of freedom for the graviton.} That being said, these considerations are in close accord with what can be realized in explicit string compactifications, and this lends significant credence to the overarching physical principles in play.

Our operating assumption in this paper will be that the cobordism conjecture is a well-motivated physical principle which is satisfied in any putative theory of quantum gravity. In other words, it imposes \textit{necessary} conditions on the
structure of a quantum gravity theory. Remarkably, this is enough to predict the existence of specific objects which trivialize the bordism group of quantum gravity. Turning the discussion around, a natural question to ask is whether the full spectrum of branes can be predicted from the triviality of the bordism group $\Omega^{\text{QG}}$.

An interesting test case is type IIB string theory since it has a non-Abelian duality group and a rich spectrum of codimension two defects known as $[p,q]$ 7-branes. The appearance of both features leads to non-trivial additional data beyond what one would expect to be captured by just a bordism group since bordism groups are generalized homology theories and as such are only sensitive to Abelian structures. As one might expect,
we find that the bordism group of $SL(2,\mathbb{Z})$ does indeed predict the appearance of D7-branes.

This begs the question as to whether we can deduce the full spectrum of $[p,q]$ 7-branes and their statistics using some mild generalization of the cobordism conjecture. We propose to accomplish this by tracking non-trivial duality twists in circle compactifications of type IIB string theory. These can be related to closed paths in the moduli space of the axio-dilaton $\mathcal{M}$. For the duality bundles on the boundary circle to trivialize in the two-dimensional bulk one needs to include physical objects of codimension two, which induce general $SL(2,\mathbb{Z})$ monodromies. These are exactly the $[p,q]$ 7-branes of the type IIB setup. In terms of $\mathcal{M}$ these physical defects are associated to a deformation of the closed paths across distinguished points given by the cusps and elliptic points of the modular curve (for $SL(2,\mathbb{Z})$ these are located at $\tau = i \infty$, $\tau = i$, and $\tau = e^{2\pi i/6}$). Tracking the non-Abelian nature of the orbifold fundamental group we can recover the full spectrum and non-Abelian statistics of the 7-branes. After the inclusion of these codimension two objects the moduli space is compactified to $\mathbb{P}^1$, for which all closed paths are contractible and the $SL(2,\mathbb{Z})$ bundles trivialize. We therefore define a generalization of the cobordism conjecture for codimension two objects predicted from the duality group acting on $\tau$ and analyze it in the context of type IIB string theory with duality group $SL(2,\mathbb{Z})$, and its non-perturbative formulation in terms of F-theory \cite{Vafa:1996xn, Morrison:1996na, Morrison:1996pp} (see e.g. \cite{Heckman:2010bq, Weigand:2010wm,
Taylor:2011wt, Weigand:2018rez} for recent reviews).

These concepts can equally well be applied to duality groups $\Gamma \subset SL(2,\mathbb{Z})$. In this case, the analogous
condition makes reference to the moduli space specified by the modular curve $\mathcal{M}_{\Gamma} = \mathbb{H} / \Gamma$ and its compactification $\overline{\mathcal{M}}_{\Gamma}$ after the inclusion of the available 7-branes, with $\mathbb{H}$ the upper half plane. For $\Gamma$ a congruence subgroup of $SL(2,\mathbb{Z})$ the corresponding compactified modular curves are Riemann surfaces. One therefore concludes that all possible $\Gamma$-bundles on the circle trivialize in the higher-dimensional bulk only if the obtained modular curve is genus zero and has no non-trivial one-cycles. This shows that the moduli space needs to be simply connected after the inclusion of the codimension two defects, in accord with the conjecture of reference \cite{Ooguri:2006in}. Moreover, we immediately deduce that there is a finite number of allowed duality groups since $\overline{\mathcal{M}}_{\Gamma}$ is only a genus zero curve for a certain choices of subgroups $\Gamma \subset SL(2,\mathbb{Z})$.

Quite remarkably, these choices do appear in the setup of F-theory compactifications in the presence of torsional sections \cite{Aspinwall:1998xj, Mayrhofer:2014opa, Hajouji:2019vxs} (see e.g. \cite{Cvetic:2018bni} for a recent review)! This data is captured by the so-called Mordell--Weil (MW) group and has direct physical implications on the global realization of the gauge group in the theory. In this sense, the presence of non-trivial MW torsion can be understood as the gauging of (part of) the center 1-form symmetries.\footnote{See \cite{Kapustin:2014gua, Gaiotto:2014kfa} for a general discussion of higher-form symmetries and their gauging, as well as \cite{Dierigl:2020myk, Apruzzi:2020zot, Bhardwaj:2020phs, Cvetic:2020kuw} for their explicit realization in terms of F-theory compactifications to six and eight dimensions.} In \cite{Cvetic:2020kuw} it was shown that the number of consistent quantum gravity theories in eight dimensions with non-simply connected gauge groups are severely restricted by a mixed anomaly involving the dynamical higher-form fields of the supergravity theory as well as the discrete center 1-form symmetries, see also \cite{Montero:2020icj} for an alternative approach leading to similar constraints. It was found that the vanishing of the anomaly has a beautiful connection to the geometry of elliptically-fibered K3 manifolds. In this way bottom up field theory arguments allow one
to exclude most of the models that cannot be derived from F-theory on an elliptic K3,
or heterotic string theory on a $T^2$, see e.g.\ \cite{Font:2020rsk}.

While some of the string theory analyses above strongly rely on the full classification of MW lattices for extremal elliptically-fibered K3 manifolds \cite{Miranda1988MordellWeilGO, QYe1999OnEE, ArtalBartolo2002MirandaPerssonsPO, shimada2000, shimada2001}, our criterion leads to a complementary viewpoint fully utilizing the presence of the duality group. Specifically, employing the refinement of the cobordism conjectures in the case of circle compactifications of type IIB we can reconstruct the full set of branes realized in the theory as well as their non-Abelian statistics under monodromy processes. For the congruence subgroups $\Gamma \subset SL(2,\mathbb{Z})$ we further find consistency constraints which restrict the presence of torsional sections. This leads to a set of possibilities slightly larger than the realized models in F-theory on elliptically-fibered K3 manifolds. The mismatch is explained by further constraints for a compact geometry and the preservation of supersymmetry. Essentially, this boils down to the further condition that we need precisely $24$ 7-branes to avoid any deficit angle in the passage from 10D to 8D vacua. So in the end we find a perfect match between the realizations of MW torsion for elliptically-fibered K3 manifolds and restrictions imposed by our proposed generalization of the cobordism conjecture.

While we have presented some extremely non-trivial checks of our proposal in the context of type IIB vacua, we expect that these considerations apply for \textit{any} quantum theory of gravity with $D$ macroscopic dimensions in which we have a non-Abelian duality group which acts on a physical parameter space and predicts the existence of codimension two defects. A particularly interesting case is that of 4D $U(1)$ gauge theory with stringlike defects. In this case, we expect that in such theories, coupling to gravity imposes strong constraints on admissible duality groups.

The rest of this paper is organized as follows. In section \ref{sec:7branes} we review the cobordism conjecture and discuss its generalizations for circle reductions of type IIB string theory. The non-Abelian structure of the defects originates from the non-Abelian $SL(2,\mathbb{Z})$ duality group. We then point out the connection to closed paths on the modular curve of the theory in section \ref{sec:modcurve}. In section \ref{sec:MW}, we extend the arguments to general congruence subgroups $\Gamma$ of $SL(2,\mathbb{Z})$. In these cases we find that the modular curve of consistent models has to be of genus zero. For compact supersymmetric configurations, we further find that the only allowed setups are realized by elliptically-fibered K3 manifolds. We conclude in section \ref{sec:concl} and present some potential avenues of further investigation. Some technical details about the derivation of the relevant bordism groups are presented in Appendix \ref{app:AHSS} and we summarize other typical examples of genus zero congruence subgroups in Appendix \ref{app:genuszero}.

\section{Bundles, Branes, and Bordisms}
\label{sec:7branes}

In this section we discuss the appearance of 7-branes in the circle compactification of type IIB string theory and its non-perturbative formulation in F-theory \cite{Vafa:1996xn, Morrison:1996na, Morrison:1996pp}. In our discussion the $SL(2,\mathbb{Z})$ duality of the type IIB string plays a central role. The duality transformation identifies different values of the fields to be physically equivalent and therefore can be understood as a redundancy of the theory. The spacetime dependent choice for the duality frame together with the transition functions therefore define an $SL(2,\mathbb{Z})$ bundle. In the following we want to explore what the presence of non-trivial duality bundles on the compactification circle imply for the existence of codimension two defects in the theory. Note that the actual duality group of type IIB string theory is given by the $Pin^+$-cover of $GL(2,\mathbb{Z})$, see \cite{Tachikawa:2018njr}. This leads to a twisting of the tangent bundle and the duality group accounting for the transformation behavior of the fermions in the system. Here, we focus on the subgroup $SL(2,\mathbb{Z})$, which is used to demonstrate the necessity to include 7-branes due to the cobordism conjecture.

Possible bundles for the group $G$ are determined by its classifying space $BG$. Since we are interested in bundles of $G = SL(2,\mathbb{Z})$ on a circle the relevant information is contained in the fundamental group of its classifying space
\begin{align}
\pi_1 \big(BSL(2,\mathbb{Z})\big) = \pi_0 \big(SL(2,\mathbb{Z})\big) \,.
\end{align}
Since $SL(2,\mathbb{Z})$ is discrete, this can be identified with the group itself. For a restriction of the duality group to a congruence subgroup $\Gamma \subset SL(2,\mathbb{Z})$ one analogously finds
\begin{align}
\pi_1 (B\Gamma) = \pi_0 (\Gamma) \simeq \Gamma \,.
\end{align}
An alternative way to understand the different bundles follows from the observation that circle bundles of discrete groups are determined by the transition function imposed when going once around the circle, i.e., by a specific group element.

\begin{figure}
	\centering
	\includegraphics[width = 0.6\textwidth]{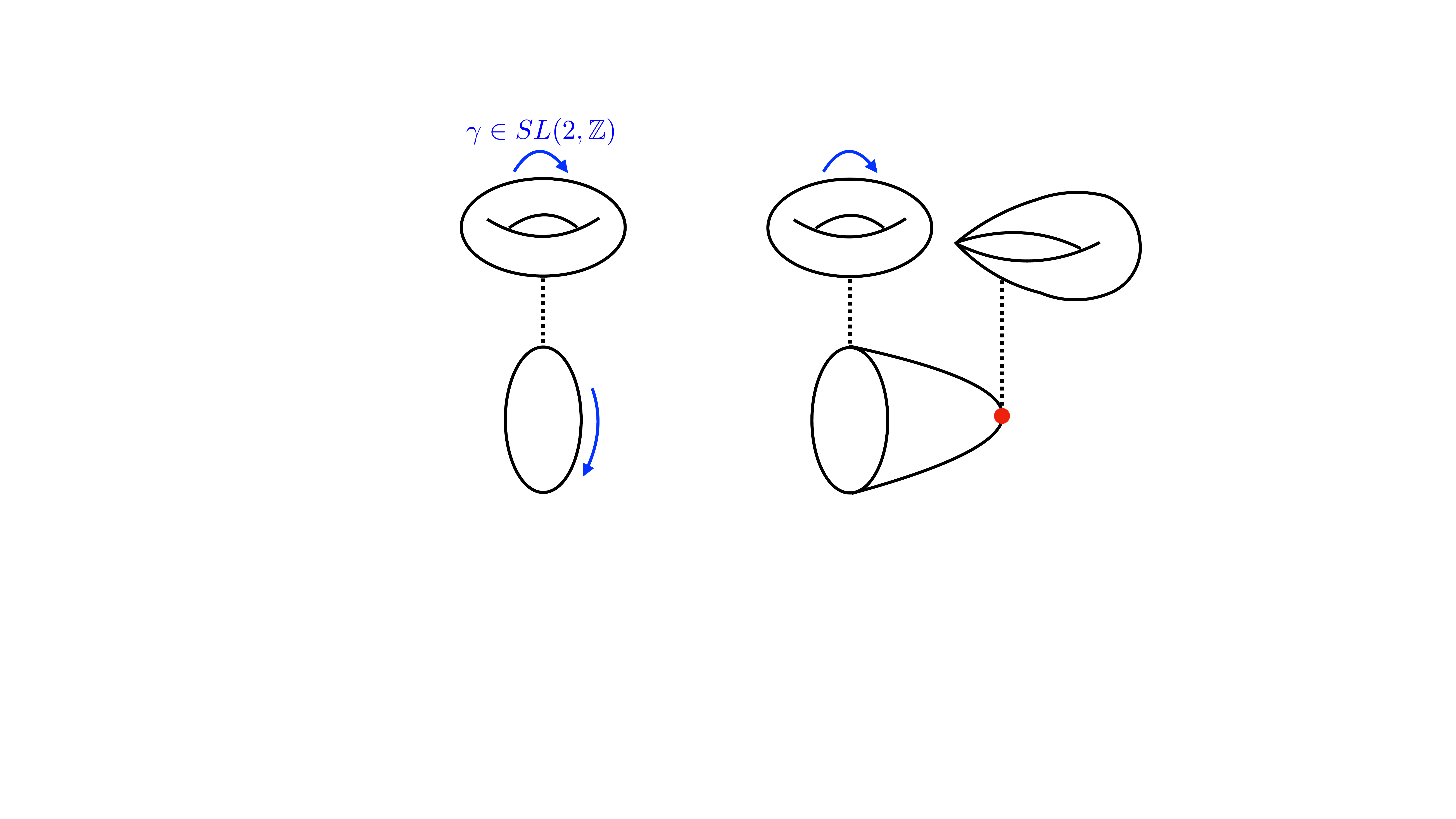}
	\caption{Left: Duality twist in the circle compactification of type IIB string theory in the F-theory description. Right: Trivialization of the bordism group by the introduction of physical objects.}
	\label{fig:circlebord}
\end{figure}

For the uniqueness of the underlying quantum theory of gravity we need to include objects that can absorb the $SL(2,\mathbb{Z})$ monodromy imposed by the non-trivial bundle. Physically, this corresponds to introducing a pointlike source in the two-dimensional geometry, or equivalently, to specifying an asymptotic profile of the $SL(2, \mathbb{Z})$ bundle along a bordant circle, as in Figure \ref{fig:circlebord}.
In type IIB string theory we know that general $[p,q]$ 7-branes will have exactly this effect. Their induced monodromy is given by
\begin{align}
\gamma_{[p,q]} = \begin{pmatrix} 1 + pq & p^2 \\ - q^2 & 1- pq \end{pmatrix} \,.
\end{align}
With the standard definition of $A$-, $B$-, and $C$-branes, see e.g.\ \cite{Weigand:2018rez}, with corresponding monodromies
\begin{align}
A: \enspace \gamma_{[1,0]} = \begin{pmatrix} 1 & 1 \\ 0 & 1 \end{pmatrix} \,, \quad B: \enspace  \gamma_{[3,1]} = \begin{pmatrix} 4 & 9 \\ -1 & -2 \end{pmatrix} \,, \quad C: \enspace \gamma_{[1,1]} = \begin{pmatrix} 2 & 1 \\ -1 & 0 \end{pmatrix} \,,
\label{eq:buildmon}
\end{align}
and the identities
\begin{align}
\gamma_{[1,0]} = T \in SL(2,\mathbb{Z}) \,, \quad \gamma_{[1,0]}^6 \gamma^{}_{[3,1]} \gamma_{[1,1]}^2 = S \in SL(2,\mathbb{Z}) \,,
\label{eq:ABC}
\end{align}
we see that we can generate each element in $SL(2,\mathbb{Z})$ by an appropriate combination of 7-branes. We further know that these objects are mutually non-local and exhibit non-Abelian statistics for monodromy processes \cite{Gaberdiel:1998mv}.

\begin{figure}
	\centering
	\includegraphics[width = 0.5\textwidth]{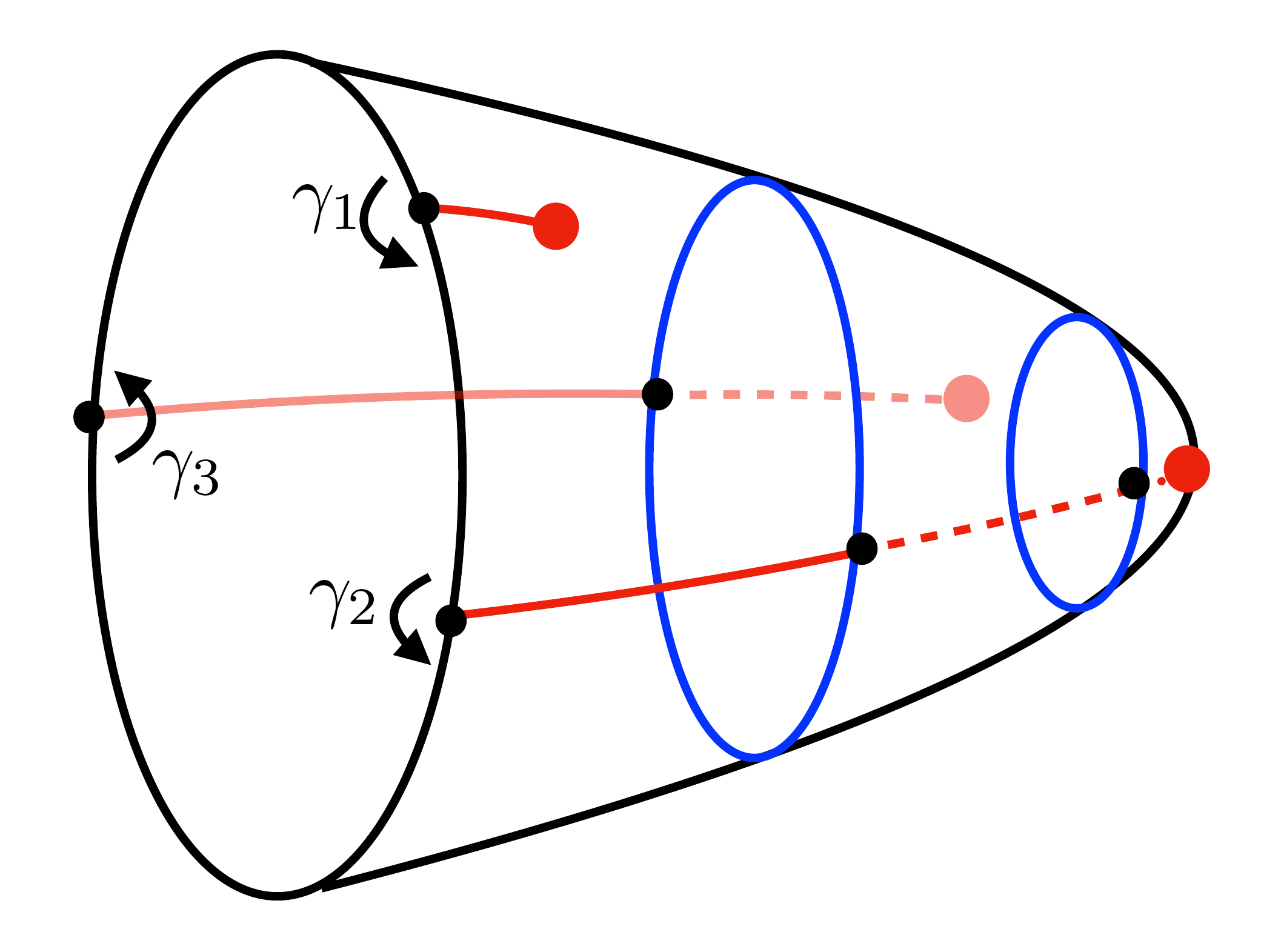}
	\caption{Decomposition of elements $\gamma \in SL(2,\mathbb{Z})$ into factors $\gamma = \gamma_1 \gamma_2 \gamma_3$ and the corresponding brane picture.}
	\label{fig:decomp}
\end{figure}

The non-Abelian statistics can be seen as follows. Starting with a general element $\gamma \in SL(2,\mathbb{Z})$ we decompose it in terms of the monodromies of the elementary building blocks \eqref{eq:buildmon} above, e.g.\ $\gamma = \gamma_1 \gamma_2 \gamma_3$ in Figure \ref{fig:decomp}. These building blocks can then be separated as individual transition functions along the circle, each of which is connected to the corresponding brane in the bulk via a branch cut.
The individual branes can then be moved around the bulk freely as long as they do not cross another branch cut. If they do cross, however, they will implement a monodromy transformation on the associated elements $\gamma_i$ while leaving the overall $\gamma$ unchanged. In this way one can reconstruct the non-Abelian statistics of the 7-branes in the theory arising from the non-Abelian monodromies.

There is also an efficient geometrical interpretation of the above discussion in F-theory as a non-perturbative description of type IIB vacua. In F-theory models the vacuum expectation value of the axio-dilaton $\tau$ is captured by the complex structure parameter of an auxiliary torus. Supersymmetry preserving F-theory vacua are then specified by genus-one-fibered Calabi-Yau manifolds. Since we are interested in type IIB compactifications on a capped off circle we focus on elliptically-fibered K3 manifolds in the F-theory framework. An elliptic K3 can be understood as a torus $T^2$ fibered over a base manifold which is given by a 2-sphere $\mathbb{P}^1$. Cutting two holes in $\mathbb{P}^1$, the base becomes topologically a cylinder, i.e.\ $S^1 \times I$ with finite interval $I$. Extending the interval to infinity this type of background can be understood as a non-perturbative circle compactification of type IIB string theory with possible duality twists as discussed above. Employing the F-theory picture, this leads to an automorphism on the elliptic fiber when going around the circle in the base, see Figure~\ref{fig:circlebord}. The closing of one side of the cylinder then leads to a singular fiber at the corresponding point if the $SL(2,\mathbb{Z})$ monodromy is non-trivial. This is a clear sign that a corresponding 7-brane has to be localized there, with monodromy action determined by the bundle.

We therefore see that the consideration of non-trivial $SL(2,\mathbb{Z})$ bundles in the circle compactification of type IIB string theory demands the introduction of general $[p,q]$ 7-branes and also imposes a non-Abelian structure when mutually non-local objects are interchanged.

We now want to compare this with the associated bordism class that takes into account the $SL(2,\mathbb{Z})$ duality structure, which for the case at hand is given by:\footnote{MD is grateful to M. Montero for enlightening discussions concerning bordism groups. See also \cite{Garcia-Etxebarria:2018ajm} for a great introduction and technical toolkit for physicists.}
\begin{align}
\Omega^{Spin}_{1} \big(BSL(2,\mathbb{Z})\big) = e(\mathbb{Z}_2, \mathbb{Z}_{12}) \,,
\label{eq:SLbord}
\end{align}
which is a group extension of $\mathbb{Z}_{12}$ by $\mathbb{Z}_2$, and necessarily Abelian. For further details on this group we refer the interested reader to Appendix \ref{app:AHSS}.

In more detail, the duality group $SL(2,\mathbb{Z})$ is generated by an element of
order four given by $S$ and an element of order six given by $R = T S$.
Note that we have $T = R S^{-1} = R S^3$. With this we can write the group $SL(2,\mathbb{Z})$ as
\begin{align}
SL(2,\mathbb{Z}) = \langle R, S: \, R^6 = S^4 = 1, \, R^3 = S^2 \rangle \,.
\end{align}
This corresponds to the structure of an amalgamated free product
\begin{align}
SL(2,\mathbb{Z}) = \mathbb{Z}_4 \ast_{\mathbb{Z}_2} \mathbb{Z}_6 \,,
\end{align}
with the non-trivial $\mathbb{Z}_2$ element embedded as $R^3$ and $S^2$ into the individual factors. Utilizing this structure one can further argue that the extension in \eqref{eq:SLbord} is trivial, see Appendix \ref{app:AHSS}.

From this representation we can also see that $S$ and $R$ respectively map to order four and six elements in the Abelianization $\mathrm{Ab}\big( SL(2,\mathbb{Z}) \big) = \mathbb{Z}_{12}$. We denote these elements as $\varphi(S)$ and $\varphi(R)$ and find
\begin{align}
\varphi(T) = \varphi(R) \varphi(S)^3 \quad \Rightarrow \quad \varphi(T)^6 = \varphi(R)^6 \varphi(S)^{18} = \varphi(S)^2 \,,
\end{align}
i.e., $T$ maps to an element of order twelve and thus necessarily generates $\mathbb{Z}_{12}$. One therefore concludes that in terms of the bordism group $\Omega_1^{Spin} \big( BSL(2,\mathbb{Z}) \big)$ the presence of the D7-brane is enough to trivialize the $\mathbb{Z}_{12}$ factor.

From the discussion above, it should be clear that the bordism group loses information about the non-Abelian properties of the involved objects. In fact, we show in Appendix \ref{app:AHSS} that for all congruence subgroups $\Gamma \subset SL(2,\mathbb{Z})$ one has
\begin{align}
\Omega^{Spin}_{1} \big(B\Gamma) = e \big(\mathbb{Z}_2, \mathrm{Ab}(\Gamma) \big) \,,
\end{align}
with $\mathrm{Ab}(\Gamma)$ the Abelianization of $\Gamma$. The existence of an associated
description in terms of a free product again suggests that the extension is trivial.

To illustrate the difference in the two approaches, let us discuss a circle compactification with duality twist given by $\gamma = S$. Since $S$ is an element of order 4 it can be represented as $e^{\pi i/2}$ in the multiplicative description of $\mathbb{Z}_{12}$. Consequently, from the bordism perspective a stack of four D7-branes would suffice in order to absorb the monodromy. However, we clearly see that $T^4 \neq S$ in terms of the full $SL(2,\mathbb{Z})$ duality group. In the alternative approach we instead see that the $S$-generator is associated to the combination of six $A$-branes, one $B$-brane, and two $C$-branes, which capture the full non-Abelian structure of the associated transition functions.

\section{Branes and Modular Curves}
\label{sec:modcurve}

In the previous section we observed that the bordism group for $SL(2,\mathbb{Z})$ does not appear to detect some of the intrinsically
non-Abelian data associated with the non-Abelian duality group. In this section we present a proposal for how to supplement this
information. In the spirit of F-theory, our main idea will be to track the profile of the supergravity field $\tau$, namely the axio-dilaton.
Our conclusions are as follows. The non-trivial $\Gamma$ bundles are translated to closed paths in the moduli space regarded as an orbifold. The inclusion of the codimension two objects, i.e., the 7-branes compactify this space to a Riemann surface. Moreover, the closed paths as elements of the orbifold fundamental group fully capture the non-Abelian statistics of the present 7-branes. If the resulting Riemann surface is of genus zero, all paths can be contracted after the inclusion of the defects and the corresponding bundles can be trivialized. If the Riemann surface is of higher genus this is not possible and not all duality bundles can be trivialized. Such duality groups can consequently not be realized in consistent theories of quantum gravity.

To frame the discussion to follow, recall that the axio-dilaton $\tau$ in F-theory compactifications takes values in the upper halfplane $\mathbb{H}$. The $SL(2,\mathbb{Z})$ duality transformation acts on it via
\begin{align}
\tau \mapsto \frac{a \tau + b}{c \tau + d} \,, \quad \text{with} \enspace \begin{pmatrix} a & b \\ c & d \end{pmatrix} \in SL(2,\mathbb{Z}) \,.
\end{align}
In this way the physically distinct values are given by values in the so-called modular curve
\begin{align}
\mathcal{M} = \mathbb{H}/SL(2,\mathbb{Z}) \,.
\end{align}
Note that only $PSL(2,\mathbb{Z})$ acts on $\tau$. However, there are other fields in type IIB string theory, such as the 2-forms $B_2$ and $C_2$ that also transform non-trivially under $\text{diag}(-1,-1) \in SL(2,\mathbb{Z})$ and the full duality group is physically meaningful.\footnote{In fact, by including the duality transformations on the gravitinos one expects that this should be extended to the metaplectic cover of $SL(2,\mathbb{Z})$, as argued in reference \cite{Pantev:2016nze}.}

We shall also be interested in the more general situation where the duality group may be a subgroup $\Gamma \subset SL(2,\mathbb{Z})$. In this case,  the same construction can be applied to duality groups which are congruence subgroups $\Gamma \subset SL(2,\mathbb{Z})$, for which one obtains the corresponding modular curves:
\begin{align}
\mathcal{M}_{\Gamma} = \mathbb{H}/\Gamma \,.
\end{align}
Including possible cusps at $i \infty \cup \mathbb{Q} \subset \mathbb{C}$ into the upper halfplane $\overline{\mathbb{H}}$ one obtains the compactification
\begin{align}
\overline{\mathcal{M}}_{\Gamma} = \overline{\mathbb{H}} / \Gamma \,,
\end{align}
which for $\Gamma$ a congruence subgroup is topologically a compact Riemann surface \cite{diamond2006first}. If there are additional elliptic points they result in orbifold points on the modular curve.

\begin{figure}
	\centering
	\includegraphics[width = 0.8\textwidth]{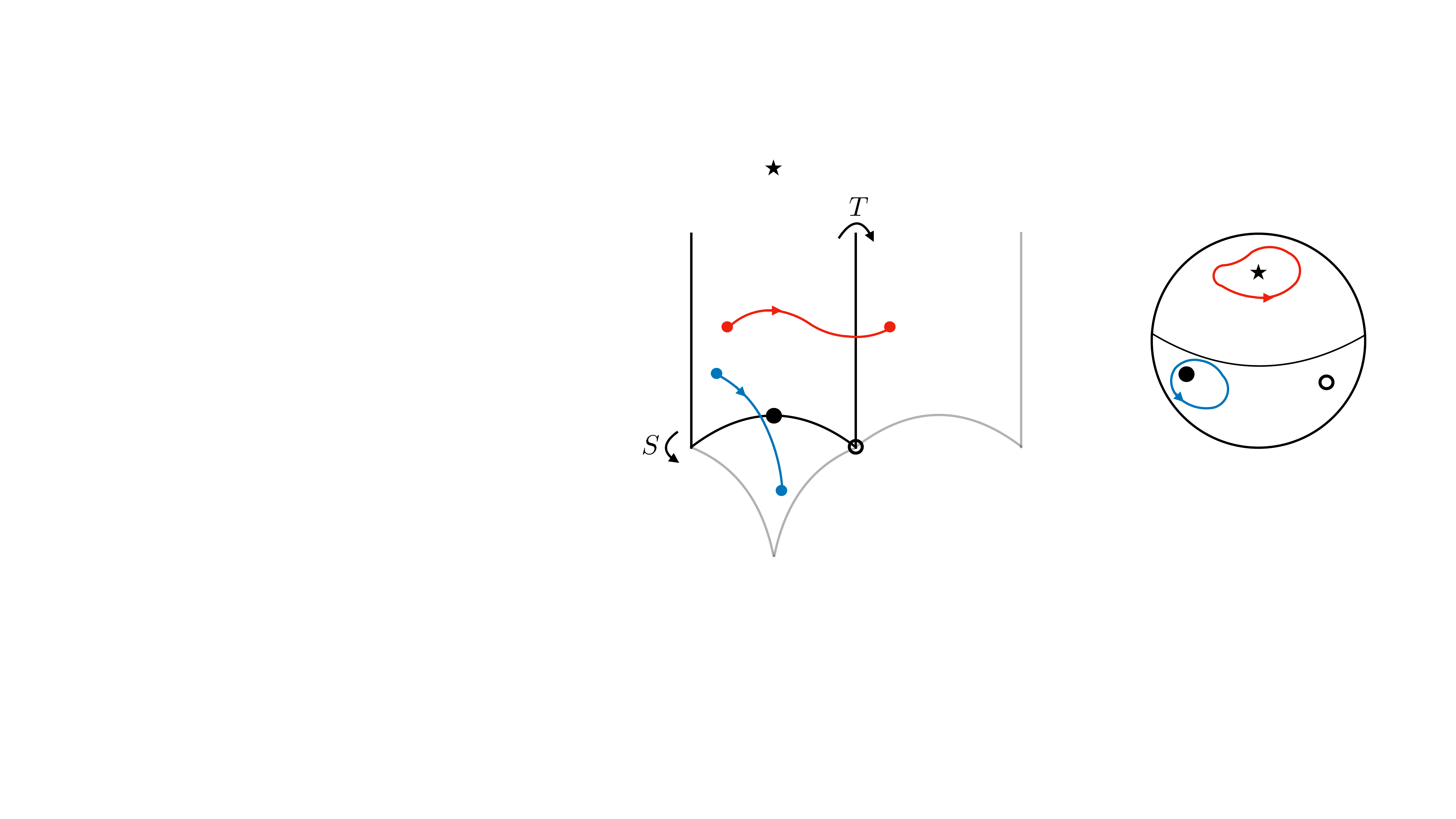}
	\caption{Two paths connecting $SL(2,\mathbb{Z})$ equivalent values of $\tau$, related by $S$ and $T$, are projected to closed paths around the cusp or one of the elliptic points in the fundamental domain.}
	\label{fig:closedpaths}
\end{figure}

We see that when going around the circle of our compactification, the axio-dilaton is identified with an $SL(2,\mathbb{Z})$ equivalent value. The corresponding map from the compactification circle to $\mathbb{H}$ therefore projects to a closed path in $\mathcal{M}$ (see Figure \ref{fig:closedpaths}), which is non-trivial for a non-trivial duality twist around the circle.
Again, the same logic can be applied for more general duality groups $\Gamma \subset SL(2,\mathbb{Z})$. In this way we can identify the closed paths on $\mathcal{M}_{\Gamma}$ with elements in $\Gamma$. This is supported by the observation that the fundamental group of $\mathbb{H}$ is trivial and therefore the orbifold fundamental group of the modular curve is given by
\begin{align}
\pi_1 (\mathcal{M}_{\Gamma}) = \Gamma \,,
\end{align}
which also takes into account the presence of possible elliptic points.
Again, note that on the level of $\tau$ one is only sensitive to the projectivization $P\Gamma$, but the inclusion of all fields in the bosonic action extends this to the full group $\Gamma$. We therefore find the important identification
\begin{align}
\pi_1 (B\Gamma) \simeq \pi_1 (\mathcal{M}_{\Gamma}) \,,
\end{align}
which relates $\Gamma$-bundles on the compactification circle with an equivalence classes of closed paths on the corresponding modular curve $\mathcal{M}_{\Gamma}$.

In the remainder of this section we want to shed light on the appearance of branes from the modular curve perspective. Let us first discuss the case of $SL(2,\mathbb{Z})$ for which the modular curve has three distinguished points. First, we have the cusp at $\tau = i \infty$ which is a marked point filled in by the compactification to $\overline{\mathcal{M}}_{\Gamma}$. This point is the stabilizer of the infinite order element $T: \tau \mapsto \tau + 1$. Additionally, we have the distinguished orbifold points $\tau = i$ and $\tau = \exp(2 \pi i / 6)$, which are fixed by finite order elements of the duality group:
\begin{align}
\tau = i \infty: \enspace T = \begin{pmatrix} 1 & 1 \\ 0 & 1\end{pmatrix} \,, \quad \tau = i: \enspace S = \begin{pmatrix} 0 & -1 \\ 1 & 0 \end{pmatrix} \,, \quad \tau = e^{2 \pi i /6}: \enspace ST = \begin{pmatrix} 0 & -1 \\ 1 & 1 \end{pmatrix} \,,
\label{eq:objmono}
\end{align}
A general closed path on $\mathcal{M}$ associated to $\gamma \in SL(2,\mathbb{Z})$ can then be composed of a concatenation of the elementary paths around two of the three distinguished points.\footnote{Choosing $\tau = i$ and $\tau = e^{2 \pi i / 6}$ the composition of closed paths directly reflects the amalgam structure $SL(2,\mathbb{Z}) \simeq \mathbb{Z}_4 \ast_{\mathbb{Z}_2} \mathbb{Z}_6$.} The extension of the bundle from the compactification circle to the bulk, as depicted in Figure \ref{fig:decomp}, can then be understood as a deformation of the closed path within the modular curve. Dragging the closed path over any of the designated points imposes the presence of an object implementing the $SL(2,\mathbb{Z})$ transformation given in \eqref{eq:objmono}, which could also be decomposed in terms of $A$-, $B$-, and $C$-branes. In this sense the elliptic points and cusps of the modular curve as an orbifold are directly related to the set of branes necessary to trivialize the configuration after the extension to the bulk. Moreover, while the fiber in the F-theory geometry becomes singular at the position of the brane, the axio-dilaton is still well-defined. Therefore, branes have to be associated to paths in the modular curve that can be contracted to one of the special points. From \eqref{eq:objmono} we see that the branes associated to the three special points are sufficient to generate the full $SL(2,\mathbb{Z})$.

Turning the logic around, starting with the orbifold modular curve we see that the inclusion of the 7-branes compactifies the curve to the topological space given by $\overline{\mathcal{M}}$. Since topologically
\begin{align}
\overline{\mathcal{M}} \simeq \mathbb{P}^1 \,,
\end{align}
one has $\pi_1 (\overline{\mathcal{M}}) = 0$ and every closed path is trivial. This means that every duality twist around the circle can be compensated by the inclusion of the corresponding branes in the bulk geometry.

For congruence subgroups one can use an equivalent approach. One starts with the orbifold modular curve with $\pi_1 (\mathcal{M}_{\Gamma}) = \Gamma$. This may contain several orbifold fixed points associated to the elliptic points of the modular curve as well as excised points that correspond to the cusps. The inclusion of the allowed 7-branes, i.e.\ the ones whose monodromy is generated by closed paths around the designated points on $\mathcal{M}_{\Gamma}$, compactifies the curve to $\overline{\mathcal{M}}_{\Gamma}$. The resulting space is topologically a Riemann surface. However, we know that $\pi_1 (\overline{\mathcal{M}}_{\Gamma}) = 0$ only if $\overline{\mathcal{M}}_{\Gamma}$ is of genus zero. This means that for congruence subgroups of higher genus the physically allowed 7-branes are not enough to allow for a trivialization of all closed paths and the associated $\Gamma$-bundles on the compactification circle. This is the case since there are closed paths that cannot be contracted to a point with a well-defined $\tau$.

At this point we make direct contact with the Swampland conjecture of reference \cite{Ooguri:2006in} which proposes that in quantum gravity, all moduli spaces are in fact simply connected. Note, however, that to apply this criterion one has to consider the compactified modular curve, that is, one has to include the physical defects that compactify the moduli space. In our context, this means we must limit our discussion to genus zero modular curves. In particular, this imposes strong constraints on the possible spectra of $[p,q]$ 7-branes compatible with a given duality group $\Gamma \subset SL(2,\mathbb{Z})$.

To test this relation we can proceed as follows. We argued that each elliptic point and cusp corresponds to a brane with a monodromy given by an element in $\Gamma$. We also noted that for genus zero congruence subgroups, these are enough to generate the full congruence subgroup, whereas for higher genus modular curves there are elements in $\Gamma$ that cannot be generated in this way. These correspond to non-trivial closed paths on $\overline{\mathcal{M}}_{\Gamma}$. This suggests a correlation between the number of generators of $\Gamma$ and the number of special points depending on the genus of the modular curve.

There are of course many choices for possible subgroups of $\Gamma \subset SL(2,\mathbb{Z})$, including those with well-appreciated connections to the theory of modular forms such as $\Gamma_{0}(k), \Gamma_{1}(k), \Gamma(k)$:
\begin{equation}
\begin{split}
\Gamma(k) &= \Big\{ \begin{pmatrix} a & b \\ c & d \end{pmatrix} \in SL(2,\mathbb{Z}): \quad \begin{pmatrix} a & b \\ c & d \end{pmatrix} = \begin{pmatrix} 1 & 0 \\ 0 & 1 \end{pmatrix} \enspace \text {mod} \enspace k \Big\} \,, \\
\Gamma_1(k) &= \Big\{ \begin{pmatrix} a & b \\ c & d \end{pmatrix} \in SL(2,\mathbb{Z}): \quad \begin{pmatrix} a & b \\ c & d \end{pmatrix} = \begin{pmatrix} 1 & \ast \\ 0 & 1 \end{pmatrix} \enspace \text {mod} \enspace k \Big\} \,, \\
\Gamma_0(k) &= \Big\{ \begin{pmatrix} a & b \\ c & d \end{pmatrix} \in SL(2,\mathbb{Z}): \quad \begin{pmatrix} a & b \\ c & d \end{pmatrix} = \begin{pmatrix} \ast & \ast \\ 0 & \ast \end{pmatrix} \enspace \text {mod} \enspace k \Big\} \,.
\end{split}
\label{eq:consub}
\end{equation}
In what follows, we shall primarily focus on the cases of $\Gamma(k)$ and $\Gamma_{1}(k)$ since these are the ones which can
lead to non-trivial Mordell-Weil torsion of the associated elliptic curve, see, however, Appendix \ref{app:genuszero} for a discussion of other genus zero subgroups.

As illustrative examples, we consider:
\begin{align}
	\begin{array}{| c | c | c | c | c |}
	\hline
	\Gamma & \text{genus} & N_g & N_p & N_g - N_p \\ \hline \hline
	SL(2,\mathbb{Z}) & 0 & 2 & 3 & -1 \\ \hline
	\Gamma_1(2) & 0 & 2 & 3 & -1 \\ \hline
	\Gamma(3) & 0 & 3 & 4 & -1 \\ \hline
	\Gamma_1 (9) & 0 & 7 & 8 & -1 \\ \hline
	\Gamma(6) & 1 & 13 & 12 & 1 \\ \hline
	\Gamma_1(11) & 1 & 11 & 10 & 1 \\ \hline
	\Gamma_1(13) & 2 & 15 & 12 & 3 \\ \hline
	\Gamma(7) & 3 & 29 & 24 & 5 \\ \hline
	\Gamma(9) & 10 & 55 & 36 & 19 \\ \hline
	\end{array}
\end{align}
where $N_g$ denotes the number of generators and $N_p$ the number of elliptic points and cusps. Accounting for the over-counting by one due to the existence of a relation between the elements, we find exactly what one would expect\footnote{Note that for $\Gamma(2)$ one finds three generators and only three special points. This is connected to the fact that one of the generators is given by $\text{diag}(-1,-1)$, which is not the case for the other subgroups.}
\begin{align}
N_g - N_p + 1 = 2 g \,,
\end{align}
associated to the non-trivial closed paths on $\overline{\mathcal{M}}_{\Gamma}$.\footnote{This is reminiscent of the standard genus formula of modular curves $g(\mathcal{M}_{\Gamma}) = 1 + \tfrac{1}{12} n - \tfrac{1}{4} N_2 - \tfrac{1}{3} N_3 - \tfrac{1}{2} N_\infty$, with $n$ the index of $\Gamma$ and $N_2$, $N_3$, and $N_\infty$ the number of elliptic points of order two, three, and cusps, respectively, i.e., $N_p = N_2 + N_3 + N_\infty$. We refer the interested reader to reference \cite{diamond2006first} for further details.}

\section{Bottom Up Bound on Mordell--Weil Torsion}
\label{sec:MW}

In the previous sections we argued that Swampland considerations lead to strong constraints on the spectrum of codimension two defects
in theories with a non-Abelian duality group $\Gamma \subset SL(2, \mathbb{Z})$. In particular, we argued that the associated modular curve $\mathcal{M}_{\Gamma}$ (as well as its compactification) must have genus zero. In this section we demonstrate that the very delicate features of Mordell--Weil torsion of an elliptically fibered K3 surface can be \textit{predicted} from Swampland considerations alone!

To frame the discussion to follow, we briefly summarize how the restriction of the duality group appears naturally in the context of F-theory with non-trivial torsional sections captured by the Mordell--Weil group, see also \cite{Aspinwall:1998xj, Mayrhofer:2014opa, Hajouji:2019vxs}. We then show that Swampland considerations correctly predict this structure. For this we start with type IIB string theory with duality group restricted to $\Gamma$ and follow the same arguments as for $SL(2,\mathbb{Z})$ above. This corresponds exactly to the setup of F-theory with non-trivial Mordell--Weil torsion. In turn this restricts some of the allowed deformations that would destroy the torsional sections and restore the full $SL(2,\mathbb{Z})$ duality structure.\footnote{We note that in the context of F-theory on an elliptically fibered K3, there are of course many complex structure moduli, and most of these are in turn specified by the ``open string moduli'' associated with the positions of
$[p,q]$ 7-branes. Tuning the positions of the 7-branes corresponds to tuning the complex structure moduli of the K3, and can produce additional structure in the elliptic fibration.}

Elliptic curves are naturally equipped with a summation law, which for the description in terms of $\mathbb{C}/\Lambda$ with two-dimensional lattice $\Lambda$ is given by the addition of complex numbers mod $\Lambda$. Applying this summation law fiber-wise this extends to a group law for sections of an elliptic fibration. The identity element is given by the zero section. Relevant for the analysis in F-theory is the set of rational sections. This property is preserved under the Abelian group law and therefore these form a group, the Mordell--Weil (MW) group. The finitely generated MW group in general takes the form:
\begin{align}
\mathrm{MW} = \mathbb{Z}^r \times \mathbb{T} \,,
\end{align}
with torsional part $\mathbb{T}$.

Now, for elliptically fibered K3 manifolds \cite{Miranda1988MordellWeilGO, QYe1999OnEE, ArtalBartolo2002MirandaPerssonsPO, shimada2000, shimada2001} as well as elliptic surfaces \cite{Miranda1986} the full set of Mordell--Weil lattices is known. In the following we will focus on the torsional part $\mathbb{T}$ for elliptically-fibered K3 manifolds. The list of allowed torsion subgroups is then given by:
\begin{equation}
\begin{split}
& \begin{array}{| c | c | c | c | c | c | c | c |}
\hline
\mathbb{T} & \mathbb{Z}_2 & \mathbb{Z}_3 & \mathbb{Z}_4 & \mathbb{Z}_5 & \mathbb{Z}_6 & \mathbb{Z}_7 & \mathbb{Z}_8  \\ \hline \hline
\Gamma & \Gamma_1(2) & \Gamma_1(3) & \Gamma_1(4) & \Gamma_1(5) & \Gamma_1(6) & \Gamma_1(7) & \Gamma_1(8) \\ \hline
\end{array} \\
& \begin{array}{| c | c | c | c | c | c |}
\hline
 \mathbb{T} & \mathbb{Z}_2 \times \mathbb{Z}_2 & \mathbb{Z}_2 \times \mathbb{Z}_4 & \mathbb{Z}_2 \times \mathbb{Z}_6 & \mathbb{Z}_3 \times \mathbb{Z}_3 & \mathbb{Z}_4 \times \mathbb{Z}_4 \\ \hline \hline
\Gamma & \Gamma(2) & \Gamma(2) \cap \Gamma_1(4) & \Gamma(2) \cap \Gamma_1(3) & \Gamma (3) & \Gamma (4) \\ \hline
\end{array}
\end{split}
\label{eq:tableMW}
\end{equation}
In the description of the fiber elliptic curve in terms of $\mathbb{C}/\Lambda$, the torsional points take a particularly simple form. For the lattice $\Lambda$ spanned by the two complex numbers $1$ and $\tau \in \mathbb{H}$ the $k$-torsion points $E(k)$ are given by:
\begin{align}
E(k) = \big\{ z \in \mathbb{C}/\Lambda: z = \tfrac{n}{k} n + \tfrac{m}{k} \tau \,, \enspace n,m \in \{0, 1, \dots , k-1\} \big\} \,.
\end{align}
For a fibration the torsional sections are associated to a fixed torsional point in each fiber. The overall complex structure $\tau$ of the fiber is free to vary. To guarantee the existence of certain torsional section one has to restrict the monodromies of the fiber along the base to leave a subset of the torsional points invariant. This reduces the full $SL(2,\mathbb{Z})$ automorphism group of the elliptic fiber to a congruence subgroup, see e.g.\ \cite{diamond2006first}. The relevant congruence subgroups for the discussion of MW torsion are $\Gamma(k)$ and $\Gamma_1 (k)$ which are defined as in \eqref{eq:consub}, for other possibilities see Appendix \ref{app:genuszero}. Elements of $SL(2,\mathbb{Z})$ not contained in the corresponding congruence subgroup act on the torsional sections and do not leave them invariant changing the geometric setup. Therefore, the restriction to the invariance of a particular subset of torsional sections enlarges the moduli space of the theory of physically inequivalent realizations to $\mathcal{M}_{\Gamma}$.\footnote{Note, however, that there are complex structure deformations that destroy the torsional sections and simultaneously enhance the duality group back to $SL(2,\mathbb{Z})$.}

The congruence subgroup $\Gamma (k)$ leaves the full set of $k$-torsion points $E(k)$ invariant, which leads to two torsional sections of the same degree, i.e., a MW torsion given by $\mathbb{Z}_k \times \mathbb{Z}_k$. The group $\Gamma_1(k)$ only preserves a single torsional section and one finds the torsion part of the MW group $\mathbb{T}$ to be $\mathbb{Z}_k$. Note, that in general one could also consider an adjoint orbit of $\Gamma_1(k)$ with respect to a coset element in $SL(2,\mathbb{Z}) / \Gamma_1(k)$. This changes the specific element in $E(k)$ which is preserved. Additionally, one has the option of two torsional sections of different degree $\mathbb{Z}_k \times \mathbb{Z}_l$. The two factors necessarily have to be of the form $l = n k$ with $n \in \mathbb{Z}$, since otherwise there would be an enhancement similar to $\mathbb{Z}_2 \times \mathbb{Z}_3 = \mathbb{Z}_6$. Thus, the corresponding congruence subgroup is given by $\Gamma(k_1) \cap \Gamma_1(k_2)$, where one can again change the $k_2$-torsional section by conjugating the $\Gamma_1(k_2)$ factor with a coset element. In the last case there are two options, either one can have $k_2 = n k_1$ in which case the torsional sections span the group $\mathbb{Z}_{k_1} \times \mathbb{Z}_{n k_1}$ or $k_1$ and $k_2$ coprime leading to $\mathbb{Z}_{k_1} \times \mathbb{Z}_{k_1 k_2}$ torsion.

Summarizing, we see that the appearance of torsional sections in elliptically-fibered manifolds is strongly correlated with the reduction of the group of automorphism of the fiber to a congruence subgroup of $SL(2,\mathbb{Z})$. Moreover, the restriction of the duality group has consequences for the moduli space of the theory given by $\overline{\mathcal{M}}_{\Gamma}$. As discussed above, the allowed compactified modular curves should be of genus zero. For $\Gamma (k)$ and $\Gamma_1 (k)$ one has \cite{cummins2003, Listgroups}:
\begin{align}
\text{genus zero:} \quad \Gamma(k) \enspace \text{with} \enspace k \in \{2, 3, 4, 5 \} \,, \quad \Gamma_1 (k) \enspace \text{with} \enspace k \in \{ 2, 3, \dots, 10, 12 \} \,.
\label{eq:set1}
\end{align}
For congruence subgroups of the form $\Gamma(k_1) \cap \Gamma(k_2)$, the ones with genus zero modular curve are given by
\begin{align}
\text{genus zero}: \quad \Gamma(2) \cap \Gamma_1(4) \,, \quad \Gamma(2) \cap \Gamma_1(3) \,, \quad \Gamma(2) \cap \Gamma_1(8) \,, \quad \Gamma(3) \cap \Gamma_1(2) \,,
\label{eq:set2}
\end{align}
which would lead to $\mathbb{Z}_2 \times \mathbb{Z}_4$, $\mathbb{Z}_2 \times \mathbb{Z}_6$, $\mathbb{Z}_2 \times \mathbb{Z}_8$, and $\mathbb{Z}_3 \times \mathbb{Z}_6$ torsion, respectively. We have also included the congruence subgroup for the realizations on K3 manifolds in the table in \eqref{eq:tableMW}. Indeed, we see that the list of genus zero subgroups contains all the allowed MW torsion groups for F-theory on K3. The torsion groups not realized for K3 surfaces are given by
\begin{align}
\mathrm{Outliers:} \,\,\, \mathbb{Z}_k \enspace \text{with} \enspace k \in \{9, 10, 12\} \,, \enspace  \mathbb{Z}_2 \times \mathbb{Z}_8 \,, \enspace \mathbb{Z}_3 \times \mathbb{Z}_6 \,, \enspace \mathbb{Z}_5 \times \mathbb{Z}_5 \,.
\end{align}
These six outlier theories are allowed from the perspective of the modular curve but seem not to be realized in 8D F-theory vacua. The reason for this is that K3 manifolds can only accommodate a total of 24 zeros of the discriminant of the elliptic fibration on the base $\mathbb{P}^1$. This is connected to the fact that the total deficit angle of the defects in the theory has to add up to $4 \pi$ in order to lead to a consistent compact geometry. However, as was shown in \cite{Hajouji:2019vxs}, K3 manifolds with MW torsion define a map from the base $\mathbb{P}^1$ in the modular curve $\overline{\mathcal{M}}_{\Gamma}$, which is a multi-cover. Therefore, the cusps appearing on the modular curve all appear as fiber singularities in the corresponding K3. When checking the outlier models above (see \cite{Listgroups}), one finds that even for a single covering one would oversaturate the bound of 24. Therefore, these MW torsion groups cannot be realized for elliptically fibered K3 manifolds and the resulting low-energy effective 8D supergravity theories.

Summarizing, we find that the generalization of the cobordism conjecture applied to a restricted monodromy group for type IIB compactifications on a circle, leads to a perfect match with the realized MW torsion for elliptically-fibered K3 manifolds. For this conclusion it was important to include the 7-brane stacks associated to the special points of the non-compactified orbifold modular curve, which trivialize the closed paths associated to non-trivial duality twists for congruence subgroups of genus zero. For higher genus modular curves the allowed 7-branes are not enough, rendering the corresponding theories inconsistent.

\section{Conclusions}
\label{sec:concl}

In this work we employed the uniqueness of quantum gravity theories in order to argue for the presence of 7-branes with non-Abelian statistics in the setup of circle compactifications of type IIB string theory. This shows that for codimension two objects the necessarily Abelian bordism groups $\Omega_1^{Spin}$ appear to not contain the full information accessible via the bundle structure. We used this to constrain the allowed MW torsion groups for F-theory compactifications to eight dimensions. This was accomplished by a restriction of the duality group to a congruence subgroup that preserves the torsional points. We also found that all duality groups of genus zero are compatible with the consistency constraints. However, some of the models prohibit a compact base manifold when preserving supersymmetry and therefore cannot be realized as an elliptically-fibered K3. Theories with a higher genus duality group are ruled out since the supersymmetric $[p,q]$ 7-branes cannot trivialize the duality twist. These conclusions have a geometric interpretation in terms of the topology of the compactified modular curves $\overline{\mathcal{M}}_{\Gamma}$. In the remainder of this section we discuss some future areas of potential investigation.

It is natural to ask if there are other more exotic objects that might ``rescue'' theories with higher genus duality groups. From what we have deduced here, such objects would need to also involve extending the field space of the theory in a way that the non-contractible cycle on the compactified modular curve becomes contractible. There is no natural candidate of such a field extension in the setup of type IIB vacua, however, one might encounter such effects in other systems with higher genus duality groups. It would be illuminating to study this possibility.

The exotic models that cannot be realized on a K3 manifold might have additional applications when considering time-dependent solutions. In fact, in \cite{Kleban:2007kk}, it was demonstrated that one can avoid the bound on the order of the discriminant when including time-dependent configurations. This can be understood as absorbing the brane induced deficit angles by a cosmological constant term. These setups clearly break supersymmetry but may lead to additional possibilities.

Further, there may be similar restrictions for duality groups in other setups. A natural testing ground is 4D $\mathcal{N}=2$ supersymmetric theories, which often possess non-trivial duality groups. We anticipate that coupling such theories to gravity might be affected by constraints similar as the ones discussed here (see also \cite{Cecotti:2018ufg, Heckman:2019bzm}). Additionally, Abelian gauge theories often transform under duality transformations and might be subject to non-trivial restrictions when coupled to gravity, see e.g., \cite{Seiberg:2018ntt, Hsieh:2019iba, Hsieh:2020jpj}. Moreover, once multiple Abelian gauge factors are considered the duality groups are enlarged to $Sp(n,\mathbb{Z})$ and one can attempt an analogous classification of the necessary physical objects for these higher-rank groups.

Much of our discussion can be extended to more general physical systems with a non-Abelian duality group and codimension two defects. In particular, we note that our considerations actually suggest a generalization to systems with a parameter space of couplings rather than a strict moduli space of vacua, as often happens in situations with stabilized moduli. In this setting, we again anticipate that there is a genus constraint on the associated modular curves.

It is also natural to consider other generalizations of bordism groups. For example, in our analysis we assumed that the
interpolating manifold had a spin structure but one might consider relaxing this to a pin structure. In the context of 4D gauge theories,
this case is of particular significance, as discussed for example in references \cite{Garcia-Etxebarria:2018ajm, Dierigl:2020wen}. Moreover, we have not considered the metaplectic cover of $SL(2,\mathbb{Z})$, which appears in the fermionic sector of type IIB string theory \cite{Pantev:2016nze, Hsieh:2019iba}. This also points towards an extension of the current approach.

In this paper we have primarily focused on lower-dimensional systems coupled to quantum gravity. On the other hand,
it is natural to ask about systems in which we take a decoupling limit where a quantum field theory
only couples to a higher-dimensional theory of gravity. This is, for example, the starting point
for most discussions of geometric engineering of quantum field theory in string theory. There is also a Swampland conjecture
that the only consistent quantum field theories are those which can be consistently coupled to a possibly higher-dimensional theory of gravity
\cite{Heckman:2018jxk}. Here we have seen some hints that the cobordism conjecture may impose non-trivial restrictions on such systems.
A further exploration of this possibility would be instructive.

The close interplay between the Swampland conjectures and
arithmetic structures such as the Mordell-Weil group of an elliptic curve hints at
a more fundamental reformulation of notions of quantum gravity in terms of
such discretized structures. It would be interesting to connect these discussions to
observations on arithmetic and moduli observed in
\cite{Moore:1998pn, Kachru:2020sio, Schimmrigk:2020dfl, Lam:2020qge, Kachru:2020abh},
as well as the more ambitious proposal of reference \cite{Samizdat}
which aims to recast some concepts from quantum fields
and strings in the framework of arithmetic geometry.
Such notions may provide a potential route for
moving the Swampland conjectures into the realm of theorems.

\section*{Acknowledgments}

We thank M. Montero, P.-K. Oehlmann, and F. Ruehle for helpful discussions.
We further thank M. Montero for comments on the manuscript.
The work of MD is supported by the individual DFG grant DI 2527/1-1.
JJH is supported by a University Research Foundation grant at the University of Pennsylvania and DOE (HEP) Award DE-SC0021484.

\newpage

\begin{appendix}

\section{Atiyah-Hirzebruch Spectral Sequence for $\Gamma \subset SL(2,\mathbb{Z})$}
\label{app:AHSS}

In this Appendix we derive the bordism groups $\Omega_1^{Spin} ( B \Gamma )$ from the Atiyah-Hirzebruch spectral sequence, see \cite{Garcia-Etxebarria:2018ajm} and references therein for a good introduction for physicists. We first briefly recall the necessary tools to evaluate the relevant bordism groups.

The spectral sequence utilizes a filtration, obtained systematically form a Serre fibration $F \rightarrow X \rightarrow B$, of the form
\begin{align}
0 = F_{-1} \Omega_n^{Spin} (X) \subset F_0 \Omega_n^{Spin} (X) \subset \dots \subset F_n \Omega_n^{Spin} (X) = \Omega_n^{Spin} (X) \,,
\end{align}
which defines the groups
\begin{align}
E^{\infty}_{k,n-k} = \frac{F_k \Omega_n^{Spin} (X)}{F_{k-1} \Omega_{n}^{Spin}(X)} \,.
\end{align}
Therefore, the bordism group can be obtained via the solution of an extension problem involving the groups $E^{\infty}_{p,q}$. For $\Omega_1$ two groups contribute and one has
\begin{align}
\Omega_1 = e (E^{\infty}_{1,0}, E^{\infty}_{0,1}) \,,
\end{align}
with the extensions defined by the short exact sequence
\begin{align}
0 \rightarrow E^{\infty}_{1,0} \rightarrow e (E^{\infty}_{1,0}, E^{\infty}_{0,1}) \rightarrow E^{\infty}_{0,1} \rightarrow 0 \,.
\end{align}
In order to obtain the groups $E^{\infty}_{p,q}$ one starts from the entries on the second sheet $E^2_{p,q}$ which are given by the homology groups
\begin{align}
E^2_{p,q} = H_p \big( B, \Omega_q^{Spin}(F) \big) \,.
\end{align}
In our case we will always use the Serre fibration with $F = pt$, from which we obtain
\begin{align}
E^2_{p,q} = H_p \big( X, \Omega_q^{Spin}(pt) \big) \,,
\end{align}
with $\Omega_0^{Spin}(pt) = \mathbb{Z}$ and $\Omega_1^{Spin}(pt) = \mathbb{Z}_2$. The third sheet is then obtained by the cohomology of the differentials
\begin{align}
d_2: \enspace E^2_{p,q} \rightarrow E^2_{p-2,q+1} \,.
\end{align}
However, since $E^{\infty}_{p,q}$ vanishes for negative $p,q$, $E^2_{1,0}$ is not affected and one has
\begin{align}
E^2_{1,0} = E^{\infty}_{1,0} \,.
\end{align}
Additionally, one has
\begin{align}
E^2_{2,0} = H_2 (X, \mathbb{Z}) \,,
\end{align}
which is trivial for the spaces we consider. Consequently, also $E^2_{0,1}$ is unaffected by the differential and we conclude
\begin{align}
\Omega_1^{Spin} (B\Gamma) = e (E^{2}_{1,0}, E^{2}_{0,1}).E
\end{align}
The involved homology groups are given by
\begin{align}
E^2_{1,0} = H_1 \big(B \Gamma, \Omega^{Spin}_{0} (pt) \big) = H_1 (B\Gamma, \mathbb{Z}) \,, \quad E^2_{0,1} = H_0 \big(B \Gamma, \Omega^{Spin}_{1} (pt) \big) = H_0 (B\Gamma, \mathbb{Z}_2)\,.
\end{align}
The first group is given by the Abelianization of $\pi_1 (B \Gamma) \simeq \pi_0 (\Gamma) \simeq \Gamma$,
\begin{align}
E^{\infty}_{1,0} = H_1 (B \Gamma, \mathbb{Z}) = \mathrm{Ab} (\Gamma) \,.
\end{align}
For $\Gamma = SL(2,\mathbb{Z})$ this is identified to be $\mathbb{Z}_{12}$. For the second involved group we note that $\Omega_1^{Spin} (pt) = \mathbb{Z}_2$, which can be understood as the two choices of the spin structure on the circle. Now, since $B \Gamma$ is path-connected we see that universally
\begin{align}
H_0 (B\Gamma, \mathbb{Z}_2) = H_0 (B\Gamma, \mathbb{Z}) \otimes \mathbb{Z}_2 = \mathbb{Z}_2 \,,
\end{align}
via the universal coefficient theorem. We therefore find that for a general congruence subgroup $\Gamma \subset SL(2,\mathbb{Z})$ one has
\begin{align}
\Omega_1^{Spin} (B\Gamma) = e \big(\mathrm{Ab}(\Gamma) , \mathbb{Z}_2 \big) \,,
\end{align}
which in particular for $SL(2,\mathbb{Z})$ leads to
\begin{align}
\Omega_1^{Spin} \big( BSL(2,\mathbb{Z}) \big) = e (\mathbb{Z}_{12}, \mathbb{Z}_{2}) \,.
\end{align}
Next, we want to argue that the extension is actually trivial using the structure of $SL(2,\mathbb{Z})$ as an amalgamated free product $\mathbb{Z}_4 \ast_{\mathbb{Z}_2} \mathbb{Z}_6$.

First, we note that the generator of $\mathbb{Z}_2$ is mapped into $\mathbb{Z}_4$ and $\mathbb{Z}_6$, by
\begin{align}
\mathbb{Z}_4: \enspace x_2 \mapsto x_4^2 \,, \quad \mathbb{Z}_6: \enspace x_2 \mapsto x_6^3 \,,
\label{eq:Z2embed}
\end{align}
where $x_i$ denotes the generator of $\mathbb{Z}_i$. Second, since bordisms are a generalized homology we have a long exact sequence of the form, see \cite{Seiberg:2018ntt},
\begin{align}
... \rightarrow \Omega_d^{Spin} (B \mathbb{Z}_2) \rightarrow \Omega_d^{Spin} (B \mathbb{Z}_4) \oplus \Omega_d^{Spin} (B \mathbb{Z}_6) \rightarrow \Omega_d^{Spin} \big( BSL(2,\mathbb{Z}) \big) \rightarrow  \Omega_{d-1}^{Spin} (B \mathbb{Z}_2) \rightarrow ...
\end{align}
With $d = 1$ one has
\begin{align}
... \rightarrow \mathbb{Z}_2 \oplus \mathbb{Z}_2 \rightarrow (\mathbb{Z}_2 \oplus \mathbb{Z}_4) \oplus (\mathbb{Z}_2 \oplus \mathbb{Z}_6) \rightarrow \Omega_1^{Spin} \big( BSL(2,\mathbb{Z}) \big) \rightarrow \mathbb{Z} \rightarrow ...
\end{align}
In all the $B\mathbb{Z}_i$ cases one can relate the first $\mathbb{Z}_2$ factor to the spin structure on the circle. The relevant bordisms groups can be found in \cite{bruner2010connective, Garcia-Etxebarria:2018ajm}, where the groups for $\mathbb{Z}_6$ can be derived from $\mathbb{Z}_2$ and $\mathbb{Z}_3$ along the lines of \cite{Seiberg:2018ntt}. Since $\Omega_1^{Spin} \big( BSL(2,\mathbb{Z}) \big)$ is torsion the last map is trivial. The identification of the first factors as the spin structure together with the embedding of $\mathbb{Z}_2$ as discussed above then specify the quotient action and suggest
\begin{align}
\Omega_1^{Spin} \big( BSL(2,\mathbb{Z}) \big) \simeq \frac{(\mathbb{Z}_2 \oplus \mathbb{Z}_4) \oplus (\mathbb{Z}_2 \oplus \mathbb{Z}_6)}{\mathbb{Z}_2 \oplus \mathbb{Z}_2} = \mathbb{Z}_2 \oplus \mathbb{Z}_{12} \,.
\end{align}
Equivalently, this suggests that
\begin{align}
\Omega_1^{Spin} (B\Gamma) \simeq \mathbb{Z}_2 \oplus \mathrm{Ab} (\Gamma) \,.
\label{eq:spinbordgamma}
\end{align}
leveraging the structure of $\Gamma$ as finite free product of cyclic groups.

In fact there is a simpler argument to show that the first spin bordism group splits\footnote{We thank the anonymous referee for pointing this out to us.}. For generalized cohomology theories $H^n (X)$ one has the splitting
\begin{align}
H^n (X) \simeq \tilde{H}^n (X) \oplus H^n (pt) \,,
\end{align}
with the reduced cohomology group $\tilde{H}^n (X)$. In the case at hand this means that $E^{\bullet}_{0,q}$ stabilizes on the second page of the AHSS and that the extension is indeed trivial. From this, \eqref{eq:spinbordgamma} follows directly.

\section{Genus Zero Congruence Subgroups}
\label{app:genuszero}
In this Appendix we discuss in greater detail congruence subgroups $\Gamma \subset SL(2,\mathbb{Z})$ associated with a modular curve of genus zero. Including $\Gamma_0(k)$ as defined in line \eqref{eq:consub}, we recall the genus zero realization in the three families of congruence subgroups $\Gamma(k)$, $\Gamma_1(k)$, and $\Gamma_0(k)$
\begin{equation}
\begin{split}
\Gamma(k):& \quad k \in \{ 2, 3, 4, 5 \} \,, \\
\Gamma_1(k):& \quad k \in \{ 2, 3, \dots, 10, 12 \} \,, \\
\Gamma_0(k):& \quad k \in \{ 2, 3, \dots, 10, 12, 13, 16, 18, 25\} \,.
\end{split}
\end{equation}
There are additional possibilities for the intersection of the groups above, e.g.\ $\Gamma_0(25) \cap \Gamma_1(5)$, and groups that cannot be described in terms of the standard congruence subgroups above. See \cite{Listgroups} for a full list and classification. These duality groups do not leave torsional sections invariant in the F-theory interpretation, but there might be other physical consequences induced by the reduction of allowed fiber automorphisms, which could be tested by explicitly constructing corresponding elliptically-fibered K3 manifolds where possible.

\end{appendix}

\bibliographystyle{utphys}
\bibliography{SwampModular}

\end{document}